# Squeeze expulsion in grain size segregation

Lu Jing and C.Y. Kwok*

*Department of Civil Engineering, The University of Hong Kong, Haking Wong Building, Pokfulam Road, Hong Kong*

Y.F. Leung

*Department of Civil & Environmental Engineering, The Hong Kong Polytechnic University, Hong Kong*

*\* corresponding author: fiona.kwok@hku.hk*

Grain segregation occurs under various conditions, such as vibration, shear and mixing. In the gravity-driven shear flow, size segregation is triggered by the percolation of small particles through the opened voids (kinetic sieving), and large particles are accumulated in the upper layer. One hypothesis for the upward migration of large particles is the squeeze expulsion mechanism, which remains yet ambiguous due to limited evidences from either physical or numerical experiments. Here we show statistically how the percolation of small particles facilitates the upward movement of large particles. We found that in large particles, the mechanical anisotropy (strong force network) coincides with the geometric anisotropy (contact network), which indicates squeeze, and the connectivity of large particles is much higher when they are squeezed through small particles. The presence of small particles filling the voids hinders the downward movements and provides dynamic 'steps' for the climbing large particles. Furthermore, increasing the coefficient of friction promotes rotation and empowers more large particles to reach the top layer, implying that the climbing particles tend to rotate relative to their neighbors. Our findings of the microstructure and movement pattern of individual particles add new evidences to the mechanism of squeeze expulsion and provide new perspective for the study of segregation.

Grain segregation is commonly found in industrial processes and in nature [1–5]. In gravity-driven shear flows, small grains preferentially fall through the local voids randomly opened by shear (i.e. kinetic sieving), and large grains drift towards the top due to imbalanced contact force (i.e. squeeze expulsion) [6]. While kinetic sieving is well established in many contexts [4,7–9], squeeze expulsion remains ambiguous as a mechanical process. Few evidences have been provided from physical experiments due to the difficulty in accessing the grain-scale information, especially the contact force, in the bulk of the flow [9–14]. On the other hand, numerical experiments, such as those by Discrete Element Modelling (DEM), can produce detailed information of velocity, volume fraction and contact force [15–20]. Statistical analysis of such information can add evidence and new insight to the mechanism of squeeze expulsion, which is the major goal of this work.

The setup of the DEM experiments is shown in Fig. 1(a). After the flow is initiated under gravity, mixing [Fig. 1(b)] and segregation [Fig. 1(c)] emerge subsequently. In Fig. 1(c), the flow reaches steady, fully developed state [15], while segregation is nearly complete. To characterize the degree of segregation, we introduce a parameter that varies from 0 (for the initial state of the bi-disperse system) to 1 (for the final state after perfect segregation). The parameter, $\alpha(t)$, is defined as $\alpha(t) = (S_0 - S(t))/(S_0 + 1)$, where $S(t)$ is the state of mixture at time $t$ compared to the initial state, i.e. $S(t) = \Delta c(t)/\Delta c_0$, and $\Delta c = c_s - c_l$ is the distance between the centers of mass of the small ($c_s$) and large particles ($c_l$) in $y$-direction. By definition, the initial condition [Fig. 1(a)] corresponds to $\alpha_0 = 0$, while $\alpha_\infty = 1$ occurs only if the two species interchange positions perfectly, i.e. $\Delta c_\infty = -\Delta c_0$. The latter case is however not possible, because the sample tends to dilate under shear [15] and diffusive mixing prevents perfect segregation [21]. Fig. 2(a) shows the evolution of $\alpha(t)$, which reveals some key features of segregation. First, segregation starts after a short period, as it takes time for the surface velocity to propagate towards the base and to generate velocity gradients. Second, the degree of segregation grows rapidly before approaching the steady state, and $\alpha(t)$ at this stage may be represented by an exponential function [19,22], with the details of the fitting shown in Fig. 2(a) insert. Third, the growth of $\alpha(t)$ plateaus almost simultaneously with the full development of steady state. This indicates that segregation is a matter of particle rearrangement due to the input of external energy, and finishes as the dynamic equilibrium is reached between the external and internal energy. Fig. 2(b) shows the concentration profiles of large particles taken at five snapshots of the segregation process.

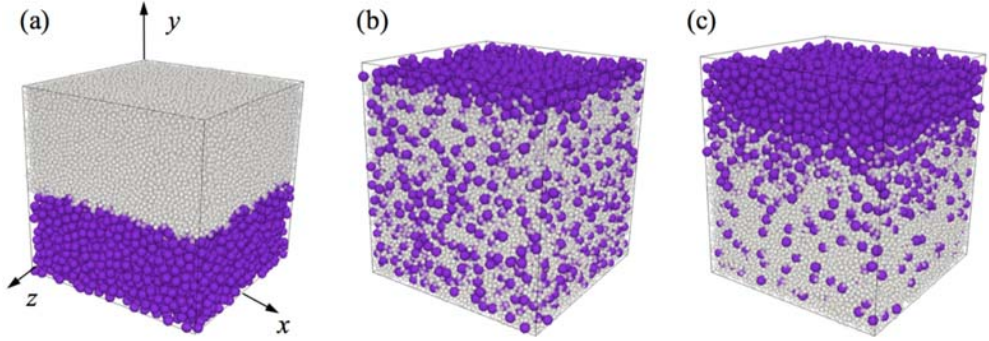

FIG. 1 (color online). Simulation set-up and different states of segregation. (a) a typical sample consists of two layers of large (violet) and small particles (grey), which are poured into the boundary box in sequence under gravity. Periodic boundaries are imposed to the flow directions ($x$) and the two sides, while the top of the box is unbounded, allowing a free flow surface. The gravity is tilted in $xy$ plane to achieve the target slope angle $\theta$. In this example, $\theta = 25°$. The diameters of large ($d_l$) and small particles ($d_s$) are 0.01 m and 0.005 m, respectively. The height of the sample is $20\,d_l$. (b) snapshot of the sample at the mixing stage. The two species are approximately uniformly mixed, with a thin layer of large particles reaching the top. (c) snapshot of the sample at the steady state. Segregation is considered to have completed. A layer of only large particles forms a cap, while a number of large particles are still immersed in the sea of small particles.

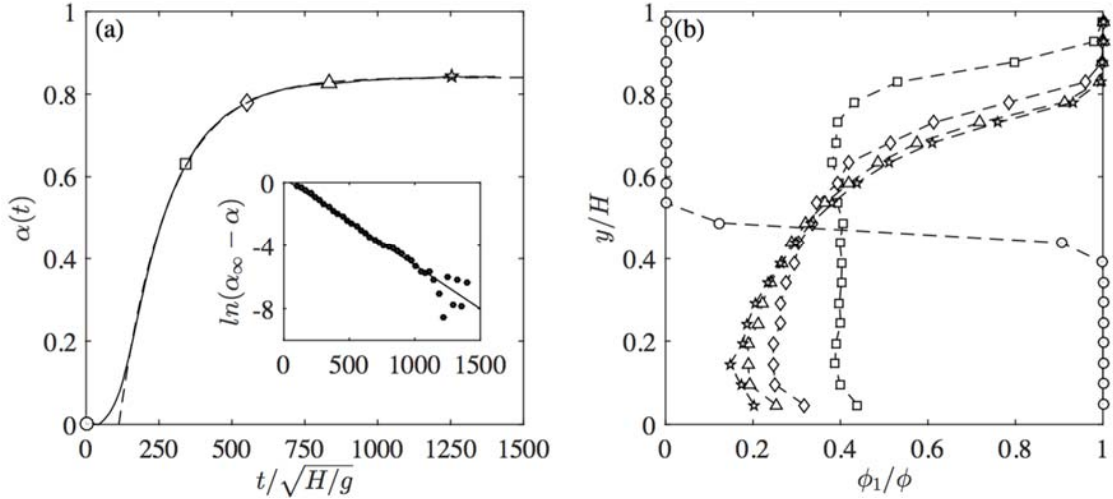

FIG. 2. Evolution of segregation. (a) the evolution of the degree of segregation, $\alpha(t)$. The definition of $\alpha(t)$ is illustrated in the text. The dash line is an exponential fitting of the solid line, given by $\alpha(t) = 0.84 - 1.64 e^{-0.042 t}$. The symbols marks $T = 0$ (○), 360 (□), 570 (◊), 850 (△) and 1250 (☆),

where $T = t/\tau$ is the dimensionless time and $\tau = \sqrt{H/g}$ is a typical time scale. The insert shows that $\ln(\alpha_\infty - \alpha) \propto 2.654 - 0.03t$, where $\alpha_\infty = 0.84$, which indicates a kinetic process of first order. (b) concentration profile of large particles, i.e. $\phi_l/\phi$, at different time. The symbols correspond to the ones specified above. Three main stages can be identified – the 'initial stage' where large particles appear only in the lower half of the sample, the 'mixing/segregation stage' where the concentration disperses more uniformly, and the 'steady stage' where the concentration profiles converge.

Next we study the movement patterns of the two species of particles. Fig. 3(a) and 3(b) show the conditional probability distributions of the final positions of particles provided their initial positions. For small particles, the different colors form roughly parallel stripes. It indicates that small particles at a certain final position have similar probabilities of coming from any initial positions, which can be attributed to the active fluctuations of small particles. In contrast, the conditional probability distribution for large particles exhibits a peak-like pattern, with a concentrated area at the top right indicated by the deep red color. This means a lower probability for large particles to migrate from the bottom to the top. The discrepant patterns show underlying differences in the microstructure of the two species, which are discussed later. The discrepancy can also be interpreted through the probability distribution of travel distances [Fig. 3(c) and 3(d)]. Contrary to the good normality for small particles (skewness $s = 0.16$), the distribution for large particles is negatively skewed ($s = -0.61$), mainly due to the higher probability of shorter (and negative) travel distances.

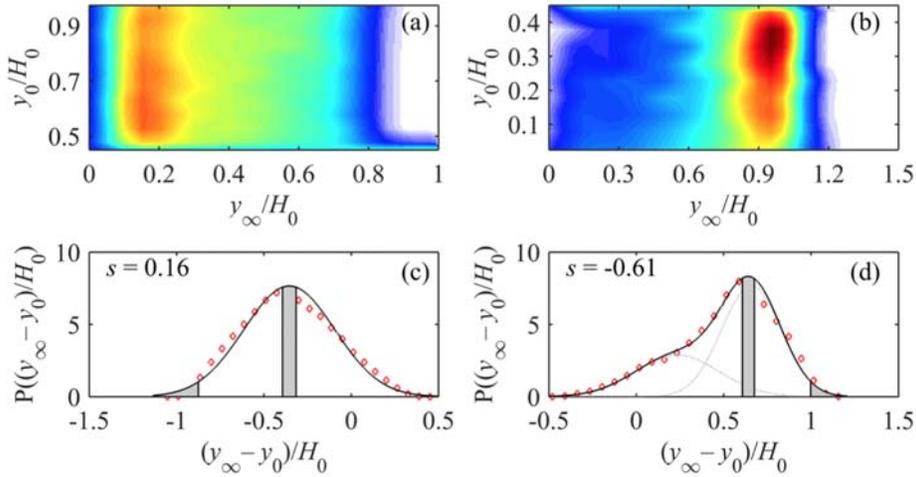

FIG. 3 (color online). Movement statistics. (a), (b), conditional probability distributions, $P(y_\infty | y_0)$,

for small and large particles, respectively. Probability gets higher as the color transitions from blue to red. The initial and final position of the particle, $y_0$ and $y_\infty$, are normalized by the initial height $H_0$. It shows that the sample height extends to $1.2 H_0$ in the steady state, and the thickness of the cap layer is ~0.4 $H_0$ (or 4 $d_l$). (c), (d), probability distributions, $P(\Delta y)$, for small and large particles, respectively, where $\Delta y = y_\infty - y_0$ is travel distance and is normalized by $H_0$. Symbols are Kernel smoothing function estimates, and $s$ denotes skewness. Black curves are fittings with normal distribution in (c) and Gaussian mixture distribution, whose two components are shown with grey dotted lines, in (d). Grey areas represent particles that travel the longest (near toe) and the most common distance (near peak) for the following analysis.

In the following we explore the microstructure, i.e. coordination number, connectivity and contact orientation, to understand the different movement patterns. As shown in Fig. 4(a) and 3(b), coordination number is as high as ~6.5 immediately after the dense sample is created. It drops as the flow gains kinetics, and bifurcates as the two species are mixed. In the mixing stage, small particles have much lower coordination number as they are accommodated in the voids through the structure of large particles, while large particles experience a dramatic increase in coordination number as small particles percolate and occupy the voids surrounding them. In the steady state, the coordination number for the two species gets closer and remains constant [20,23]. Based on these observations, a reasonable hypothesis is that segregation correlates with connectivity [6,21,24], which is confirmed in the particles with the longest travel distance. In particular, less velocity fluctuations can be found as the small particles are sinking through the middle under gravity [Fig. 4(c)]. The very low connectivity allows a higher probability for them to find voids underneath. After they reach the bottom layer, they are subjected to intrinsic velocity fluctuations as in shear flows [23,25,26]. For the large particles [Fig. 4(d)], despite the velocity fluctuations on the top (in the steady state), they move rapidly through the middle of the sample, with mostly positive upward velocity and significantly higher connectivity. Note that the pattern of upward velocity is less significant when all particles are under examination [grey points in Fig. 4(c) and 4(d)]. It is attributed to the fact that segregation is a matter of fluctuation and intermittence, and the temporal-spatial averaged behavior of the entire species tends to be trivial. Since the rapid migration is only identified in the particles with the longest travel distance, it is logical to propose that the necessary condition for large particles to successfully migrate from the bottom to the cap is that they continuously move up when heavily crowded and expulsed by the small particles.

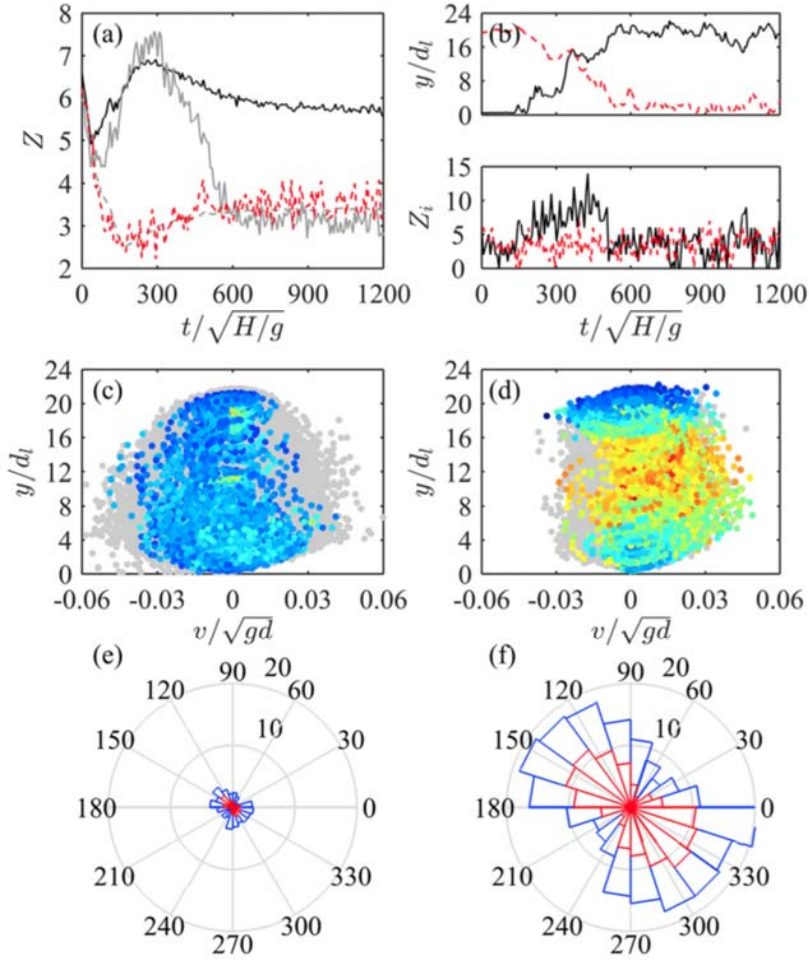

FIG. 4 (color online). Microstructure. (a) Coordination number $Z$ plotted against non-dimensional time $t/\sqrt{H/g}$, where $Z = \sum Z_i / N$, in which $Z_i$ is connectivity (the number of contact) of each particle, and $N$ is the number of particles. Light (grey) solid and dashed lines are for all large and small particles, respectively. Dark (blue) solid and (red) dashed lines are for the large and small particles with the longest travel distance. (b) normalized elevation $y/d_l$ and $Z_i$ plotted against $t/\sqrt{H/g}$, for one single large (solid lines) and small (dashed lines) particle, both of which travelled the longest distance in *y*-direction. When the large particle is migrating through the mixture, $Z_i$ increases to 12, while for the dropping small particle, it oscillates around 4. (c), (d), normalized upward velocity $v/\sqrt{gd_l}$ of the small and large particles, respectively, as they appear at different elevations $y/d_l$ in the sample. The grey points are for the particles with the most common travel distance, while the colored points are for the ones that travel the longest distance, as discussed in Fig.

3. The colors are scaled by connectivity $Z_i$, where red is the highest (~12) and blue the lowest (0) value of $Z_i$. (e), (f), rose diagrams for the small and large particles, respectively, with the longest travel distance. Blue is the contact network, while red is the strong force (i.e. $F_n > 1$) network. $F_n$ is the normal contact force normalized by its mean (i.e. $F_n = f_n / \langle f_n \rangle$).

In Fig. 4(e) and 4(f), two sets of polar distribution are presented; one is the contact network responsible for the geometric anisotropy, and the other is the strong force network responsible for the mechanical anisotropy [12]. Generally, the contact orientation is consistent with the studies of mono-disperse flows [23,27], as it is determined mainly by the slope angle. The size of the rose diagram is what distinguishes the two species in the bi-disperse system. The small particles are more free to propagate in any directions due to the small size of contact network. This finding bridges microstructure to kinetic sieving [6] and explains the movement pattern shown in Fig. 3(a) and 3(c). For the large particles, the great number of contacts implies a strong barrier against their movements in the contact orientation (geometric anisotropy). Since the mechanical anisotropy (i.e. strong force network) coincides with the geometric anisotropy, it indicates that the most probable direction of expulsion is normal to the strong force. This normality agrees with the definition of 'squeeze', despite the fact that it can occur either upward (30–60°) or downward (210–240°). Once a large particle is expulsed to a higher layer, the small particles dropping into the voids may prevent its settling to the previous position. It is therefore statistically probable that a number of large particles continuously climb up, while some other particles stay in the lower layers due to fluctuations, hence the movement pattern shown in Fig. 3(b) and 3(d).

Since the contact network is proved critical, it is natural to further investigate the role of friction in segregation. Interestingly, both previous studies [19] and our simulations (results omitted for brevity) reveal that neither slope angle (as long as steady state is achieved) nor the proportion of large particles (in a moderate range) changes $\alpha_\infty$. In contrast, we find that the coefficient of friction, $\mu$, has a significant influence on $\alpha_\infty$. Increasing $\mu$ empowers more large particles to reach the top [Fig. 5(a) and 5(b)]. In Fig. 5(a), nearly all cases approach steady state within approximately the same time, as the segregation speed is determined mainly by the slope angle [19]. Associated with friction are the sliding and rotation at contact, since $\mu$ is essentially the upper limit of the ratio of tangential force to normal force (i.e. $|f_t| \leq \mu f_n$), and sliding occurs once the limit is reached [25]. It is found that more sliding occurs in lower friction cases [Fig. 5(c)], where the small particles can percolate more easily under gravity, and $\alpha_\infty$ is negatively correlated with sliding fraction [Fig. 5(d)]. On the other hand, Fig. 5(e) shows that in higher friction cases, it is more probable for large particles to rotate related to their neighbors [see Fig. 5(e) legend], and $\alpha_\infty$ is positively correlated with $\sigma_{PDF}$ [Fig. 5(f)]. We thus propose that apart from squeeze expulsions, the relative rotation between particles also promotes the climbing action of large particles in the segregation process.

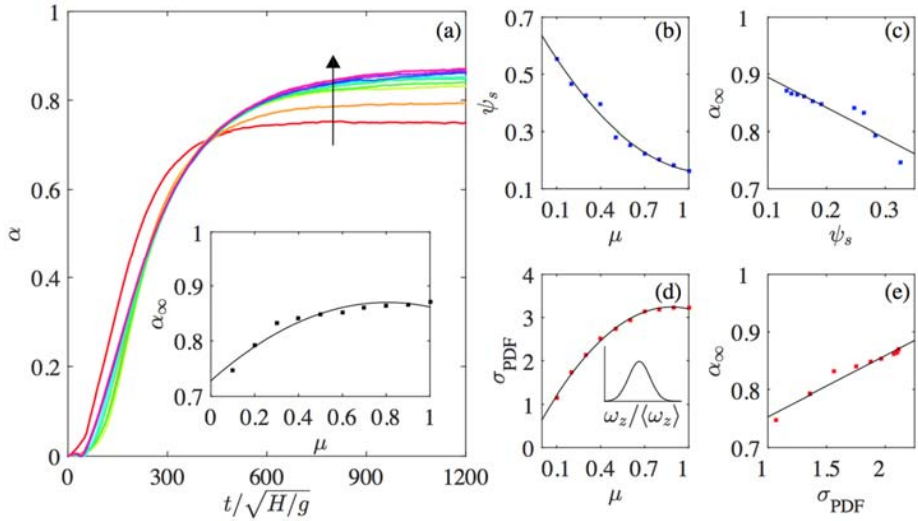

FIG. 5 (color online). Role of friction in segregation. (a) evolution of $\alpha$ with varying $\mu$ for the given slope angle $\theta = 25°$, where $\mu$ is varied from 0.1 to 1.0. Arrow points to higher $\mu$. Inset: $\alpha_\infty$ as an increasing function of $\mu$. (b) sliding fraction, $\psi_s$, as a function of $\mu$. (c) correlation between $\alpha_\infty$ and $\psi_s$. (d) the standard deviation of rotational velocity, $\sigma_{PDF}$, as a function of $\mu$. The rotational velocity is normalized by its mean in each layer, i.e. $\omega_z/\langle\omega_z\rangle$, and it obeys normal distribution (insert). The mean $\mu_{PDF}$ is $-1$ for all cases, and higher $\sigma_{PDF}$ indicates more relative rotation between particles and their neighbors. Statistically, it means a higher probability for particles to climb up. (e) correlation between $\alpha_\infty$ and $\sigma_{PDF}$.

This work interprets the percolation and expulsion in segregation with the statistics of contact network and movement pattern, which sheds light on the study of mechanism in such granular patterns as mixing, segregation and stratification. It is also in line with attempts towards a theoretical framework of granular segregation, where the specified percolation velocity is responsible for segregation and diffusive mixing is incorporated [11,18,21,24]. A recent experimental study [11] showed that the percolations of large and small particles are essentially different. This, together with findings of the current study, such as the different movement patterns and the role of friction, may be incorporated into the framework [28]. Moreover, although the theoretical consideration of diffusive mixing avoids perfect segregation by smearing the interface [21,24], the reasons for large particles being left within the sea of small particles (in the steady state) will be further investigated.

Looking towards the future, we will explore some well-recognized but yet unexplained issues in segregation, such as the onset of segregation with increasing shear rate and the unaltered final degree of segregation with different slope angles [19]. These issues, which can be studied from the microscopic perspective, are vital to the understanding of macroscopic granular patterns.

The work was supported by Research Grants Council of Hong Kong (under RGC/GRF 17203614), and the Research Institute for Sustainable Urban Development at The Hong Kong Polytechnic University. The computation was performed using the HKU Information Technology Services research computing facilities that are supported in part by the Hong Kong UGC Special Equipment Grant (SEG HKU09).